\documentclass[conference]{IEEEtran}
\IEEEoverridecommandlockouts
\usepackage{cite}
\usepackage{amsmath,amssymb,amsfonts}
\usepackage{algorithmic}
\usepackage{graphicx}
\usepackage{textcomp}
\usepackage{xcolor}
\usepackage{booktabs}
\usepackage{subcaption}
\usepackage{multirow}
\usepackage{tikz}

\def\BibTeX{{\rm B\kern-.05em{\sc i\kern-.025em b}\kern-.08em
    T\kern-.1667em\lower.7ex\hbox{E}\kern-.125emX}}

\usepackage[
  per-mode=symbol,
  per-symbol=/,
  detect-all
]{siunitx}

\newcommand\copyrighttext{%
  \footnotesize This work has been submitted to the IEEE for possible publication. Copyright may be transferred without notice, after which this version may no longer be accessible.}
\newcommand\copyrightnotice{%
\begin{tikzpicture}[remember picture,overlay]
\node[anchor=south,yshift=11pt] at (current page.south) {\fbox{\parbox{\dimexpr\textwidth-\fboxsep-\fboxrule\relax}{\copyrighttext}}};
\end{tikzpicture}%
}
    
\begin{document}

\title{
New Number Formats for FFT IP Cores in Optical OFDM Transceivers\thanks{This paper was partly funded by the German Federal Ministry of Research, Technology and Space as part of the project Open6GHub+ (16KIS2406).}
}

\author{\IEEEauthorblockN{Lukas Krupp\IEEEauthorrefmark{1}, Gustavo Magalhães Gomes De Souza\IEEEauthorrefmark{1}, Sani Nassif\IEEEauthorrefmark{2} and Norbert Wehn\IEEEauthorrefmark{1}}
\IEEEauthorblockA{\IEEEauthorrefmark{1}RPTU University Kaiserslautern-Landau, Kaiserslautern, Germany\\
\IEEEauthorrefmark{2}Radyalis LLC, Austin, TX, USA}
}

\maketitle
\IEEEpubid{\begin{minipage}{\textwidth}
  \copyrightnotice
\end{minipage}} 

\begin{abstract}
Many state-of-the-art digital signal processing (DSP) implementations use fixed-point arithmetic due to its reduced hardware complexity and high throughput compared to conventional floating-point arithmetic. In contrast, machine learning accelerators exhibit substantial gains from reduced-precision floating-point formats, enabling improvements in energy efficiency and peak throughput. These advances motivate a re-evaluation of numerical representations for classical DSP workloads. A central question is whether reduced-precision floating-point formats can achieve competitive power, performance, and area compared to fixed-point implementations, while providing advantages in dynamic range and numerical robustness.
This paper presents a new cross-layer co-design methodology for DSP kernels that jointly optimizes numerical representations, arithmetic units, and application-level performance. As a case study, we focus on the fast Fourier transform (FFT), a fundamental DSP block across many applications. The FFT is evaluated within optical orthogonal frequency-division multiplexing (OFDM) transceivers, where it dominates power consumption and silicon area as FFT size and modulation order scale to support data rates beyond \SI{100}{\giga\bit\per\second}. We compare fixed-point and reduced-precision floating-point formats using post-layout power and area results in a \SI{12}{\nano\meter} FinFET technology and demonstrate system-level performance in terms of bit error rate (BER) versus $\mathbf{E_b/N_0}$.
For a 256-point FFT engine in a \SI{128}{\giga\bit\per\second} transceiver, we show that 11- and 12-bit custom floating-point formats preserve BER performance close to a 32-bit floating-point reference across multiple modulation orders, while reducing FFT core power by up to 19.8\% and area by up to 12.0\% compared to representative fixed-point designs. To the best of our knowledge, this is the first investigation of custom reduced-precision floating-point arithmetic for FFT cores in optical OFDM transceivers.
\end{abstract}


\section{Introduction}
Many state-of-the-art digital signal processing (DSP) implementations rely on fixed-point arithmetic, enabling low hardware complexity, reduced power consumption, high throughput, and deterministic behavior compared to floating-point arithmetic. Consequently, fixed-point formats have long been the preferred choice for performance-critical kernels in communication systems, multimedia, and other DSP applications.

Recent advances in machine learning (ML) accelerators have demonstrated that reduced-precision floating-point arithmetic can deliver dramatic improvements in energy efficiency and computational throughput. Over the past decade, the performance of ML accelerators has increased by nearly two orders of magnitude every 24 months \cite{b1}. The most significant gains originate from architectural and numerical innovations. Aggressive reduction of floating-point precision has proven to be a key enabler. These techniques have seen widespread adoption both in industry, e.g., by GPU architectures from NVIDIA \cite{b2}, and in academia, e.g., within the open-source RISC-V architectures from the PULP Platform \cite{b3}.

These developments motivate a re-evaluation of numerical representations for classical DSP workloads. Floating-point formats inherently provide a larger dynamic range and improved numerical robustness. However, it remains open whether reduced-precision floating-point arithmetic can achieve competitive power consumption, silicon area, and throughput compared to fixed-point implementations. The key question addressed in this work is therefore whether the efficiency gains achieved with reduced-precision floating-point arithmetic in ML can be transferred to DSP kernels.

To answer this question, we adopt a cross-layer perspective that links numerical representation, hardware implementation, and system-level performance. As a case study, we focus on the fast Fourier transform (FFT), a foundational DSP kernel required in a wide range of DSP applications. The FFT is evaluated in the context of optical orthogonal frequency-division multiplexing (OFDM) transceivers, where it represents one of the dominant contributors to power consumption and silicon area \cite{b4}. The use case is highly relevant for next-generation data-center interconnects and high-speed serializer–deserializer (SerDes) links, where optical OFDM enables high spectral efficiency under strict energy-per-bit constraints \cite{b5}. 
In this paper, we make the following key contributions:
\begin{itemize}
    \item A value-distribution-based optimization framework that systematically derives application-aware reduced-precision number formats for DSP kernels.
    \item A detailed design-space exploration of an FFT IP core targeting a \SI{128}{\giga\bit\per\second} optical OFDM transceiver, including post-layout power and area analysis using a state-of-the-art \SI{12}{\nano\meter} FinFET technology.
    \item A system-level validation within the OFDM transceiver, demonstrating that custom reduced-precision floating-point architectures outperform fixed-point implementations in combined accuracy, power, and area efficiency.
\end{itemize}

To the best of our knowledge, this is the first investigation of custom reduced-precision floating-point arithmetic for FFT cores in optical OFDM transceivers. 

\section{Background and Related Work}
\label{sec:background}

\subsection{Optical OFDM Transceivers}
The relevance of optical OFDM transceivers is strongly increasing due to current ML trends. In the era of large language models (LLMs), the rapid growth of model sizes has led to exponentially increasing demands on both inference and training performance. Two fundamental scaling strategies exist to meet these demands: \emph{scale-up}, which increases the performance of individual compute nodes, and \emph{scale-out}, which distributes workloads across massively parallel nodes. As single-node scaling struggles to keep pace with model growth, scale-out has become a necessity. Data centers deploy hundreds of thousands of interconnected accelerators \cite{b6}.

This architectural shift places high pressure on data-center interconnect bandwidth. Optical links, and in particular co-packaged optics (CPO), have emerged as state-of-the-art solutions due to their superior bandwidth density and energy efficiency in terms of \si{\pico\joule\per\bit} compared to electrical interconnects \cite{b7}. The bottleneck in such systems lies in the electrical front-end, in particular in the SerDes pipelines \cite{b5}.

While current high-speed optical links \cite{b8} rely on non-return-to-zero (NRZ) and pulse-amplitude modulation (PAM-4), OFDM is explored as a promising candidate for next-generation optical SerDes. OFDM enables data rates beyond \SI{200}{\giga\bit\per\second} per wavelength through high spectral efficiency and improved robustness against channel impairments. However, these advantages come at the cost of DSP complexity \cite{b4}. In optical OFDM transceivers, FFT and IFFT blocks dominate the digital power and silicon area as FFT sizes and modulation orders scale to achieve peak throughputs. Consequently, FFT IP cores become a key determinant of the achievable energy efficiency and throughput of optical OFDM systems.

\begin{figure}[htbp!]
\centerline{\includegraphics[width=0.45\textwidth]{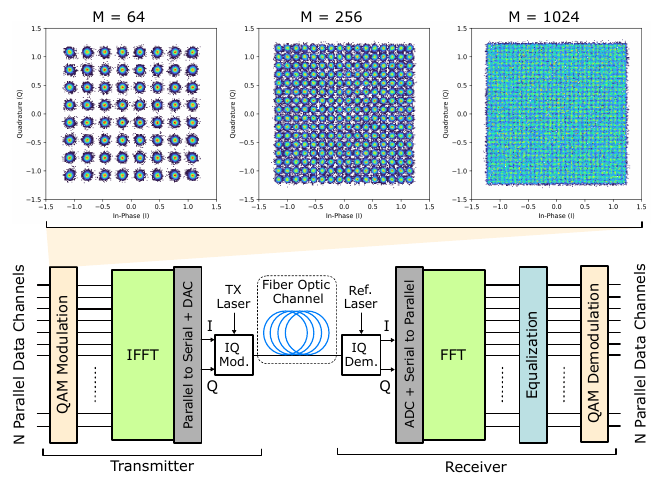}}
\caption{Optical OFDM transceiver employing $M$-QAM modulation with $N$ parallel data channels.}
\label{fig:ofdm_qam}
\end{figure}

Fig.~\ref{fig:ofdm_qam} shows the architecture of an optical OFDM transceiver. To increase spectral efficiency and throughput, optical OFDM systems scale along two dimensions: the QAM modulation order $M$ and the FFT size $N$. Increasing $M$ densifies the signal constellation, as shown in Fig.~\ref{fig:ofdm_qam}, reducing the minimum Euclidean distance between constellation points. This increases sensitivity to noise and tightens the signal-to-noise ratio (SNR) requirements. Higher-order QAM imposes stricter precision requirements on the underlying DSP blocks.

Increasing the FFT size $N$ enables finer subcarrier granularity and higher degrees of parallelism, thereby supporting higher aggregate data rates. However, larger FFTs also increase the internal dynamic range due to signal accumulation across multiple stages and raise the number of arithmetic operations. As a result, FFT blocks dominate the power consumption and silicon area of the digital baseband as $N$ scales. In high-throughput coherent optical links, the joint scaling of $M$ and $N$ is a key driver of both performance and implementation cost. Due to the resulting relevance of dynamic range and numerical precision, application-specific number formats emerge as a natural design lever to balance power, area, and numerical robustness in next-generation optical OFDM transceivers.

\subsection{Related Work}
The interest in custom number formats has been driven by deep neural networks (DNNs). The transition from IEEE-754 floating-point to reduced-precision floating-point formats (e.g., BF16, FP8, or FP4) has demonstrated substantial improvements in energy efficiency and throughput while maintaining model accuracy \cite{b9}. In the context of DNNs, automated optimization flows for custom floating-point systems have been proposed \cite{b10}. However, DNN workloads are inherently tolerant to quantization noise, unlike classical DSP applications, where numerical errors directly affect signal integrity.

Custom floating-point arithmetic has further been explored in high-level synthesis (HLS) frameworks targeting FPGA implementations \cite{b11}. While such approaches demonstrate architectural flexibility and resource savings, also within selected applications \cite{b12}, system-level performance metrics are not incorporated into the precision selection process and accelerators are treated as stand-alone components.

In FFT-based OFDM transceivers, prior work has focused on word-length optimization to balance quantization noise and hardware cost \cite{b13}. Adaptive-precision FFT architectures have been proposed to reduce energy consumption under BER constraints. However, these approaches rely on fixed-point arithmetic and do not investigate reduced-precision floating-point formats.
To the best of our knowledge, this work is the first to investigate custom reduced-precision floating-point formats in optical OFDM transceivers, jointly considering hardware implementation cost and communication performance.

\section{Methodology}
\label{sec:method}
Fig.~\ref{fig:method} shows the proposed cross-layer co-design methodology. The flow starts from a DSP application like optical OFDM and its specification. The parameters define the operating regime of the signal processing chain and directly influence the numerical requirements imposed on its DSP kernels.

\subsection{Overview}
\label{subsec:overview}

\begin{figure*}[htbp]
  \centering
  \includegraphics[width=0.8\linewidth]{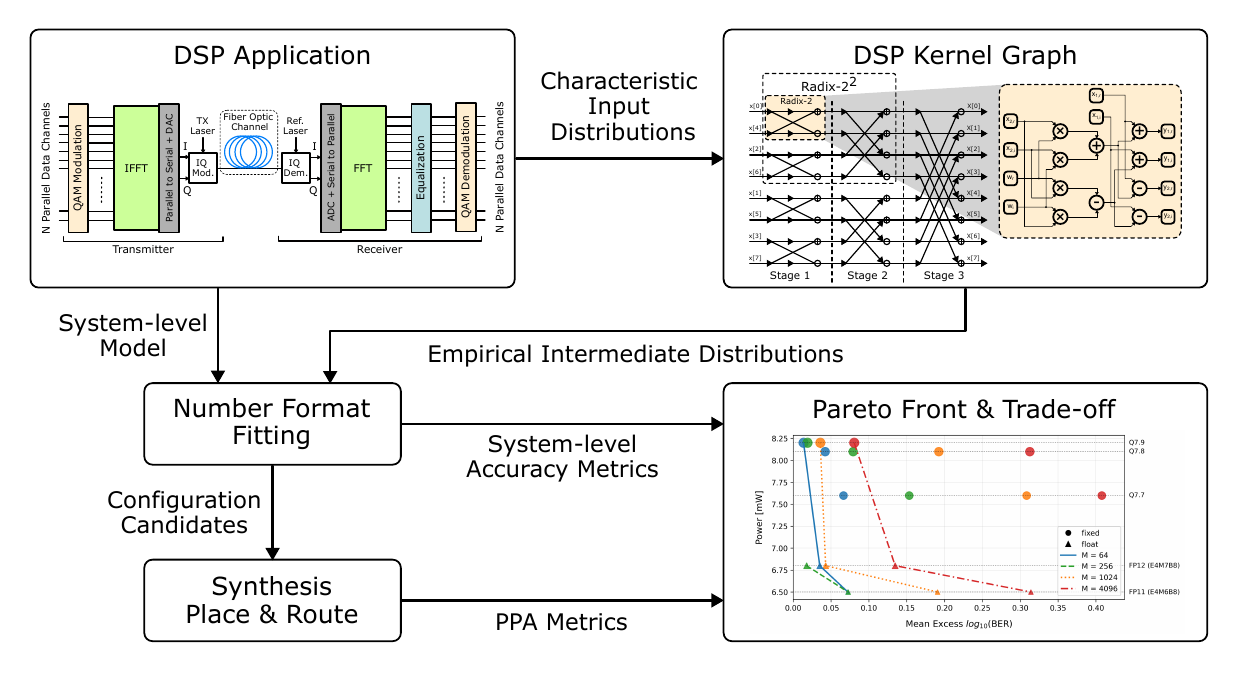}
  \caption{Cross-layer co-design methodology for number formats in digital signal processing systems.}
  \label{fig:method}
\end{figure*}

The methodology first represents the DSP kernels under investigation, such as FFT or IFFT blocks, as executable directed acyclic graphs (DAGs) and embeds these models into a system-level simulation of the target DSP application. Representative input signal distributions are propagated through the DAG to obtain empirical value distributions and dynamic ranges at the internal nodes. These distributions are then used to fit candidate number formats to individual operations or groups of operations, after which the corresponding arithmetic units are instantiated in register-transfer level (RTL) microarchitectures. The resulting designs are finally evaluated with respect to application-level accuracy and implementation efficiency, quantified through post-place-and-route (P\&R) power, performance, and area (PPA) analysis. In this work, the FFT is used as a representative DSP kernel. The FFT is modeled as a composition of radix-2 butterfly units, as shown for an eight-input example in Fig.~\ref{fig:method}. Each butterfly is expressed as a DAG of real-valued arithmetic operations.

Since multiple number formats may satisfy the system-level accuracy constraints, the approach yields a set of design points. The candidates form a Pareto front capturing the trade-offs between implementation efficiency and numerical accuracy. 

\subsection{Distribution-based Number Format Fitting}
Based on the described DAG representation, number format selection is driven by empirical value distributions at internal nodes of the DSP kernel. Given a set of representative input vectors extracted from the application-level model, the DAG is executed and intermediate values are recorded at each node. Let $\mathcal{D}_v$ denote the empirical value distribution observed at node $v$.
To characterize the numerical range requirements, multiple statistical descriptors are extracted from $\mathcal{D}_v$. These include tail statistics derived from selected percentiles (e.g., upper quartiles) and the maximum observed absolute values. 

The custom reduced-precision floating-point format consists of one sign bit, $b_e$ exponent bits, $b_m$ mantissa bits, and an explicit exponent bias $B$, resulting in a real value
\begin{equation}
x = (-1)^S (1 + m) 2^{e - B},
\end{equation}
where $e$ denotes the stored exponent value and $m$ the fractional
mantissa value.
The exponent bit width $b_e$ determines the dynamic range, while the mantissa bit width $b_m$ controls the precision. The bias $B$ shifts the representable range. A custom floating-point format is denoted as FP$\langle 1+b_e+b_m\rangle$ with $\text{E}\langle b_e\rangle \text{M}\langle b_m\rangle \text{B}\langle B\rangle$.
Unlike standard floating-point, the parameters $(b_e, b_m, B)$ are treated as tunable variables and are aligned to the empirical node distributions $\mathcal{D}_v$.
To reduce hardware complexity, features associated with IEEE-compliant floating-point like exception handling are omitted. 

Number format fitting is performed using an iterative scheme. The exponent width, mantissa width, and bias are swept. For each candidate format, the representable range is compared against the statistical descriptors of the corresponding DAG node or node group distributions.

\section{Experimental Setup and Implementation}
\label{sec:setup}
We implement a graph-based framework in C++ that models DSP kernels as executable DAGs. Using this library, FFT computation graphs are constructed. The FFT-DAG is wrapped as a Python function that mirrors the interface of the standard \textit{NumPy} FFT. The function accepts complex-valued input vectors and the format configuration \(F\), which specifies the numerical representation evaluated.

\subsection{Optical OFDM Transceiver Simulation} 

The FFT and IFFT implementations are integrated into a coherent optical OFDM transceiver simulation built on the \textit{OptiCommPy} Python library \cite{b14}. A high-throughput optical OFDM configuration is evaluated, as summarized in Table~\ref{tab:exp_setup}. The data rate is fixed to \SI{128}{\giga\bit\per\second}, while the modulation order \(M\) is varied across a wide range to reflect increasing spectral-efficiency targets. This setup enables a systematic study of tightening SNR and quantization-noise requirements as both modulation order and internal FFT dynamic range increase.

\begin{table}[h]
\centering
\caption{Experimental setup and simulation parameters}
\label{tab:exp_setup}
\begin{tabular}{l l}
\toprule
\textbf{Parameter} & \textbf{Value} \\
\midrule
Data rate & \SI{128}{\giga\bit\per\second} \\
FFT size $N$ & 256 \\
QAM orders $M$ & 64, 256, 1024, 4096 \\
Channel model & AWGN \\
Channel estimation & Pilot-based, frequency-domain \\
Equalization & Single-tap per subcarrier (ZF) \\
Number of OFDM frames & 1000 \\
Metric & BER vs.\ $E_b/N_0$\\
\bottomrule
\end{tabular}
\end{table}

\subsection{FFT Microarchitecture}
The FFT hardware is implemented as a parametrizable IP core based on a single-path delay feedback (SDF) radix-$2^2$ pipeline architecture. The SDF-FFT architecture \cite{b15} is widely adopted and provides a controlled microarchitectural baseline to compare arithmetic formats.
For fixed-point designs, dynamic-range growth is controlled by stage-wise scaling, i.e., division by two after the butterfly stages. This state-of-the-art technique avoids the accumulation of integer bits. Fixed-point operands use signed two's-complement, where $b_i$ denotes the number of integer bits and $b_f$ the number of fractional bits. The total word length is $W = b_i + b_f$. For a given input distribution, $b_i$ is fitted so that changing $W$ only changes fractional precision. We denote fixed-point configurations as INT$\langle W\rangle$. Floating-point designs use the same format throughout the pipeline and do not employ scaling. The per-stage bit width assignment is not yet considered.

All hardware variants are synthesized with the same target clock frequency of \SI{500}{\mega\hertz}. Achieving the full system-level throughput of an optical OFDM transceiver may require parallel SDF-FFT lanes. Therefore, the reported results focus on relative area, power, and accuracy trends between arithmetic formats under the same architecture, rather than claiming that a single SDF lane alone realizes the complete transceiver rate.

\subsection{Accuracy and Hardware Metrics}
\label{sec:metrics}
For all candidate FFT cores synthesis and P\&R are performed in a \SI{12}{\nano\meter} FinFET technology using an industrial design flow based on Synopsys \textit{Design Compiler} and \textit{IC Compiler~II}. For each configuration, we report post-P\&R silicon area as well as the estimated total power consumption. To capture the system-level impact of numerical distortions, BER-based accuracy metrics are adopted as the primary criterion for evaluating the numerical robustness of the FFT cores.

For each arithmetic configuration and modulation order $M$, simulations of the OFDM transceiver are performed, and the resulting BER curves are evaluated over a sweep of SNR measured as $E_b/N_0$. All BER results are evaluated relative to a 32-bit floating-point reference implementation.
Let $\mathrm{BER}_{\mathrm{F}}(\mathrm{SNR}_i)$ denote the BER obtained with a given arithmetic format at SNR point $\mathrm{SNR}_i$, and let $\mathrm{BER}_{\mathrm{T}}(\mathrm{SNR}_i)$ denote the target BER with 32-bit floating-point precision. The excess logarithmic BER at each operating point is defined as
\begin{equation}
\Delta_i = \log_{10}\!\left(\mathrm{BER}_{\mathrm{F}}(\mathrm{SNR}_i)\right)
           - \log_{10}\!\left(\mathrm{BER}_{\mathrm{T}}(\mathrm{SNR}_i)\right).
\end{equation}

This logarithmic formulation reflects the exponential sensitivity of BER in practical systems and allows aggregation across operating points. Positive values of $\Delta_i$ indicate a degradation relative to the reference, while $\Delta_i = 0$ corresponds to identical performance.
From the set of SNR values where both target and reduced-precision BER are non-zero, the mean ($\mu_{\Delta}$) and the maximum ($\Delta_{\max}$) excess logarithmic BER are extracted. They capture the typical performance loss across the operating region and the worst-case behavior. 

\section{Results}
\label{sec:results}
We first analyze the stage-wise unscaled value distributions of the FFT under representative optical OFDM input signals. At the FFT input, the signal magnitudes are clustered around unity due to the QAM constellation in the OFDM transmitter, as shown in Fig.~\ref{fig:ofdm_qam}. As the signal propagates through the FFT, the dynamic range increases gradually across stages, as expected due to additions in the butterflies.
A key outcome of the fitting process is that the required dynamic range fits within a custom floating-point format with $b_e=4$ and $B=8$. 

\subsection{System-Level Accuracy}
\label{sec:accuracy}
\begin{figure*}[t]
  \centering

  \begin{subfigure}{0.47\textwidth}
    \centering
    \includegraphics[width=\linewidth]{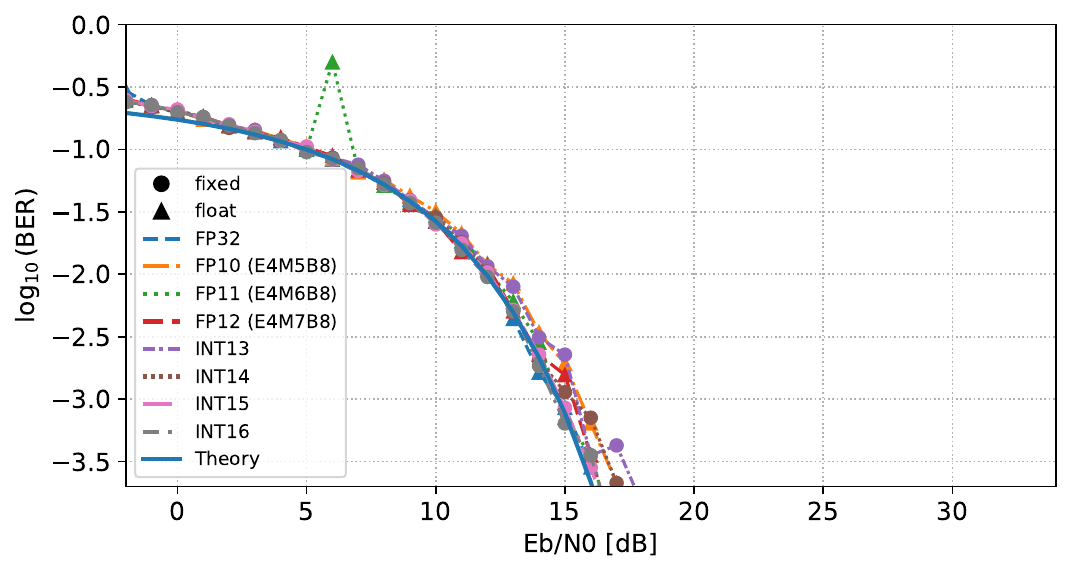}
    \caption{64-QAM}
    \label{fig:ber_64qam}
  \end{subfigure}
  \hfill
  \begin{subfigure}{0.47\textwidth}
    \centering
    \includegraphics[width=\linewidth]{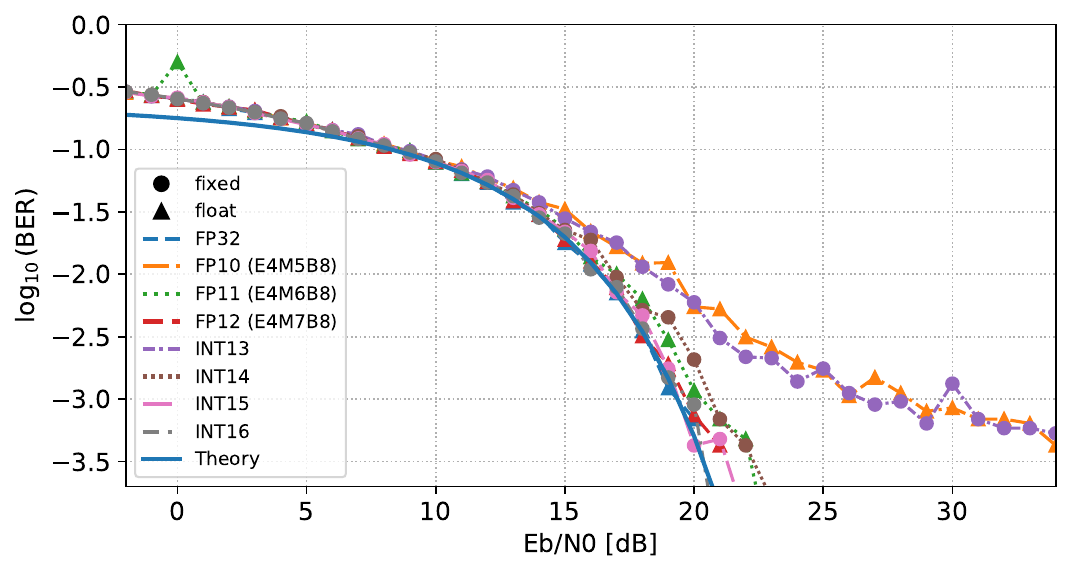}
    \caption{256-QAM}
    \label{fig:ber_256qam}
  \end{subfigure}

  \medskip

  \begin{subfigure}{0.47\textwidth}
    \centering
    \includegraphics[width=\linewidth]{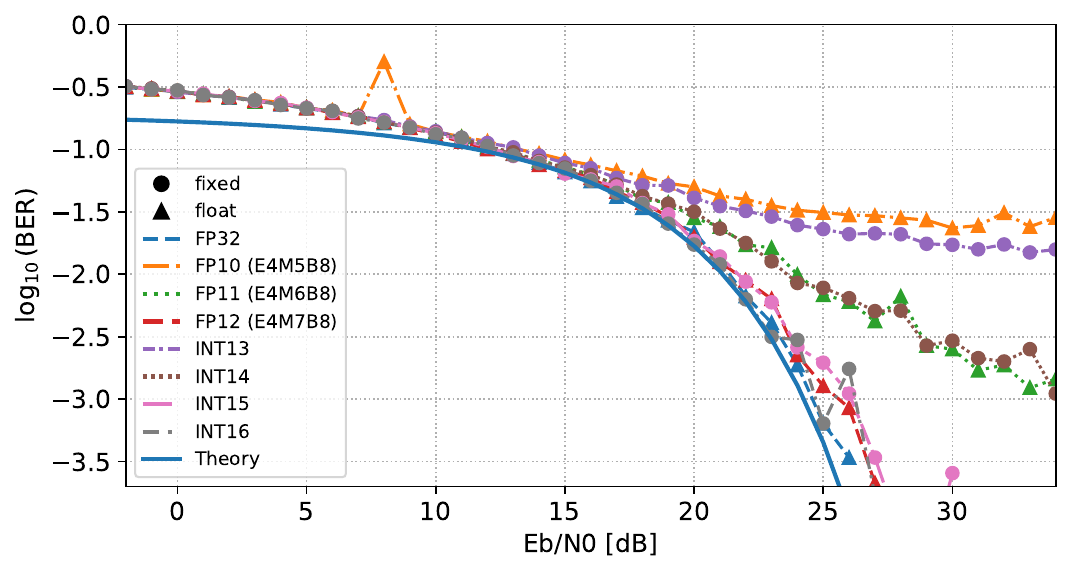}
    \caption{1024-QAM}
    \label{fig:ber_1024qam}
  \end{subfigure}
  \hfill
  \begin{subfigure}{0.47\textwidth}
    \centering
    \includegraphics[width=\linewidth]{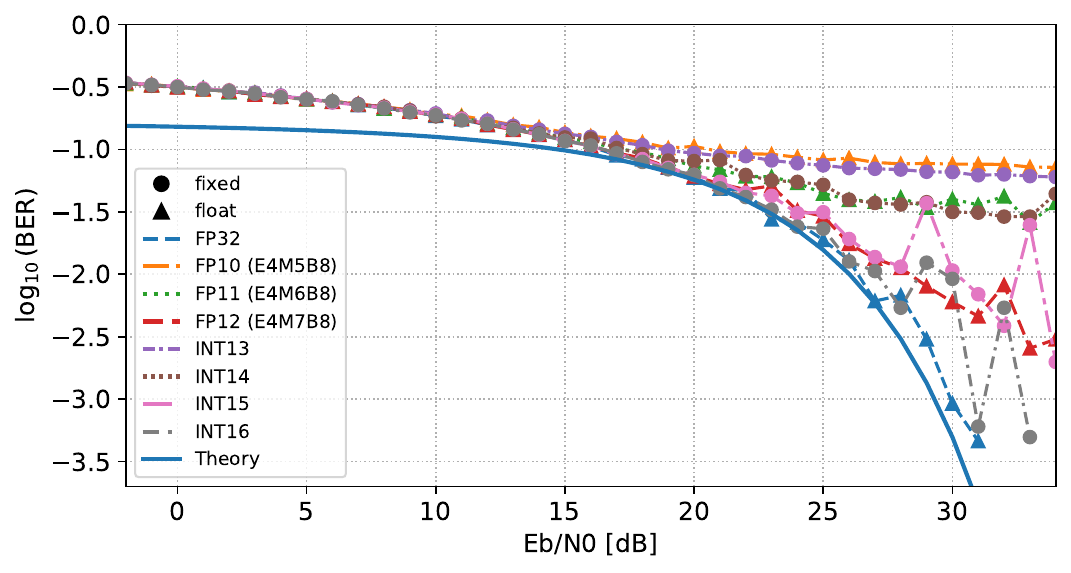}
    \caption{4096-QAM}
    \label{fig:ber_4096qam}
  \end{subfigure}

  \caption{BER versus $E_b/N_0$ for increasing QAM modulation orders comparing FP32, reduced-precision floating-point, and fixed-point FFT/IFFT implementations.
  }
  \label{fig:ber_qam_scaling}
\end{figure*}

Once the exponent range covers the observed value growth, the remaining bit budget can be allocated to the mantissa ($b_m$) to meet the precision requirements of the application.
Therefore, we evaluate the BER performance of the optical OFDM transceiver for multiple QAM modulation orders using different numerical representations within the FFT.
Fig.~\ref{fig:ber_qam_scaling} shows BER curves as a function of $E_b/N_0$ for 64-QAM to 4096-QAM. Results are reported for a full-precision FP32 FFT as reference, several reduced-precision floating-point candidates (FP10, FP11, and FP12 with $b_e=4$ and $B=8$), and a set of fixed-point implementations (INT13 to INT16).

For reduced-precision floating-point formats, BER performance depends primarily on the available mantissa width. For lower-order modulations such as 64-QAM, even aggressive formats (e.g., FP10) achieve BER curves nearly indistinguishable from FP32 over the evaluated $E_b/N_0$ range. As the modulation order increases, constellation points become denser and the system becomes increasingly sensitive to numerical noise introduced by reduced-precision arithmetic.
This trend is particularly evident for 1024-QAM and 4096-QAM, where insufficient mantissa precision leads to a pronounced BER floor at moderate to high $E_b/N_0$. Among the evaluated floating-point candidates, FP12 consistently tracks the FP32 reference across all QAM orders, while FP11 represents a marginal but still acceptable configuration depending on the target BER. FP10 exhibits early saturation for higher modulation orders.

Fixed-point implementations show a similar behavior, with BER performance improving monotonically as fractional precision increases. However, fixed-point formats require more fractional bits to approach the floating-point baseline.

\subsection{Hardware Results and Trade-Off}
\label{sec:ppa}

Table~\ref{tab:ppa_ber} reports post-P\&R power and area for the fixed-point and floating-point FFT cores implemented in a \SI{12}{\nano\meter} FinFET technology. In addition, the table includes the application-level accuracy metrics introduced in Section~\ref{sec:metrics}. We omit configurations FP10 (E4M5B8) and INT13, which are shown in Fig.~\ref{fig:ber_qam_scaling}, since their numerical robustness degrades rapidly.

\begin{table*}[!t]
\centering
\vspace{1mm}
\caption{Post-P\&R power and area in \SI{12}{\nano\meter} FinFET and numerical robustness metrics for FFT core variants. Reported values show mean and maximum excess $\log_{10}(\mathrm{BER})$ relative to floating-point reference for different modulation orders $M$.}
\label{tab:ppa_ber}
\setlength{\tabcolsep}{5.0pt}
\renewcommand{\arraystretch}{1.1}
\begin{tabular}{@{}lccccccccccc@{}}
\toprule
\multirow{2}{*}{Format} &
\multirow{2}{*}{Power (\si{\milli\watt})} &
\multirow{2}{*}{Area (\si{\milli\meter\squared})} &
\multicolumn{2}{c}{$M{=}64$} &
\multicolumn{2}{c}{$M{=}256$} &
\multicolumn{2}{c}{$M{=}1024$} &
\multicolumn{2}{c}{$M{=}4096$} \\
\cmidrule(lr){4-5}\cmidrule(lr){6-7}\cmidrule(lr){8-9}\cmidrule(lr){10-11}
 &  &  &
$\mu_{\Delta}$ & $\Delta_{\max}$ &
$\mu_{\Delta}$ & $\Delta_{\max}$ &
$\mu_{\Delta}$ & $\Delta_{\max}$ &
$\mu_{\Delta}$ & $\Delta_{\max}$ \\
\midrule
INT14 & 7.6 & 0.023 &
0.0665 & 0.4260 &
0.1532 & 0.9353 &
0.3084 & 1.7889 &
0.4077 & 2.1288 \\
INT15 & 8.1 & 0.025 &
0.0421 & 0.3979 &
0.0793 & 0.5677 &
0.1924 & 1.2759 &
0.3125 & 1.8296 \\
INT16 & 8.2 & 0.027 &
0.0138 & 0.0969 &
0.0190 & 0.1165 &
0.0357 & 0.7097 &
0.0807 & 0.9983 \\
\midrule
FP11 (E4M6B8) & 6.5 & 0.022 &
0.0723 & 0.7792 &
0.0725 & 0.3865 &
0.1908 & 1.2492 &
0.3141 & 1.8921 \\
FP12 (E4M7B8) & 6.8 & 0.025 &
0.0351 & 0.2632 &
0.0178 & 0.1946 &
0.0430 & 0.3979 &
0.1348 & 1.0000 \\
\bottomrule
\end{tabular}
\end{table*}

\begin{figure*}[h]
  \centering

  \begin{subfigure}[t]{0.47\textwidth}
    \centering
    \includegraphics[width=\linewidth]{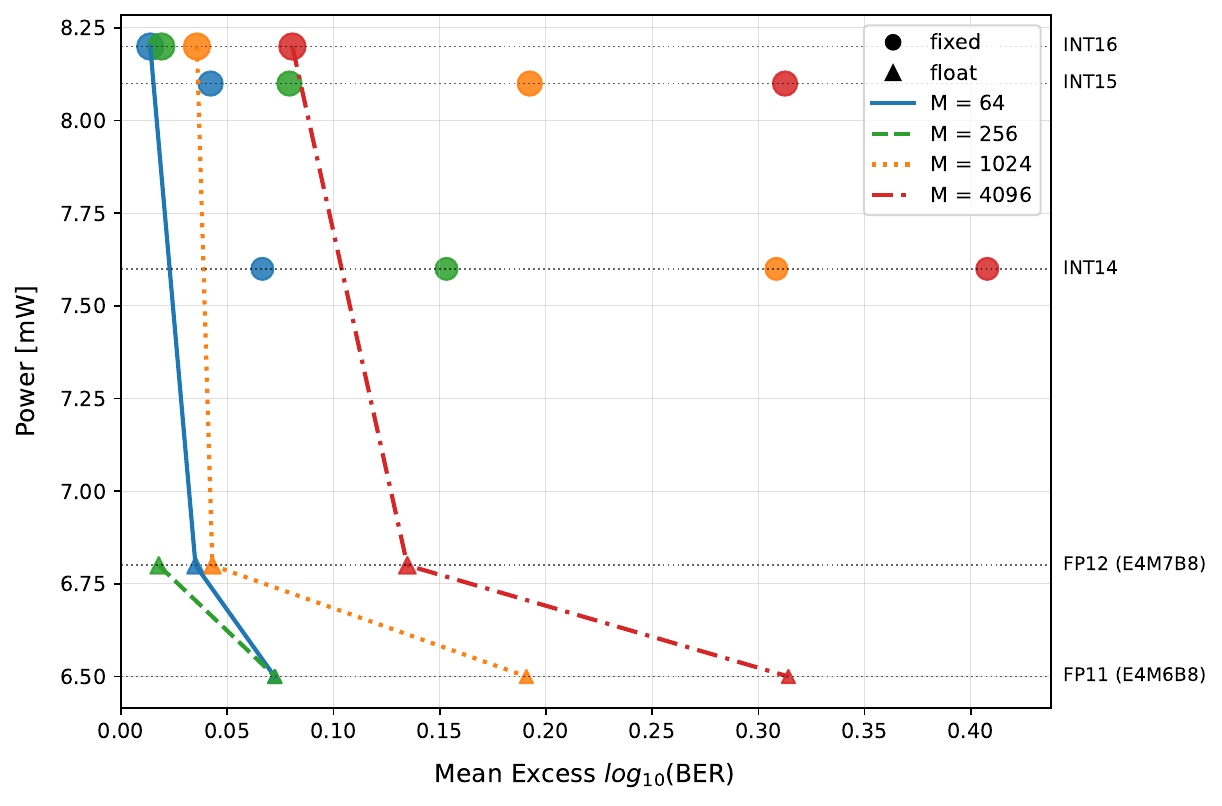}
    \caption{Accuracy--power per-$M$ Pareto fronts.}
    \label{fig:pareto_acc_power}
  \end{subfigure}\hfill
  \begin{subfigure}[t]{0.47\textwidth}
    \centering
    \includegraphics[width=\linewidth]{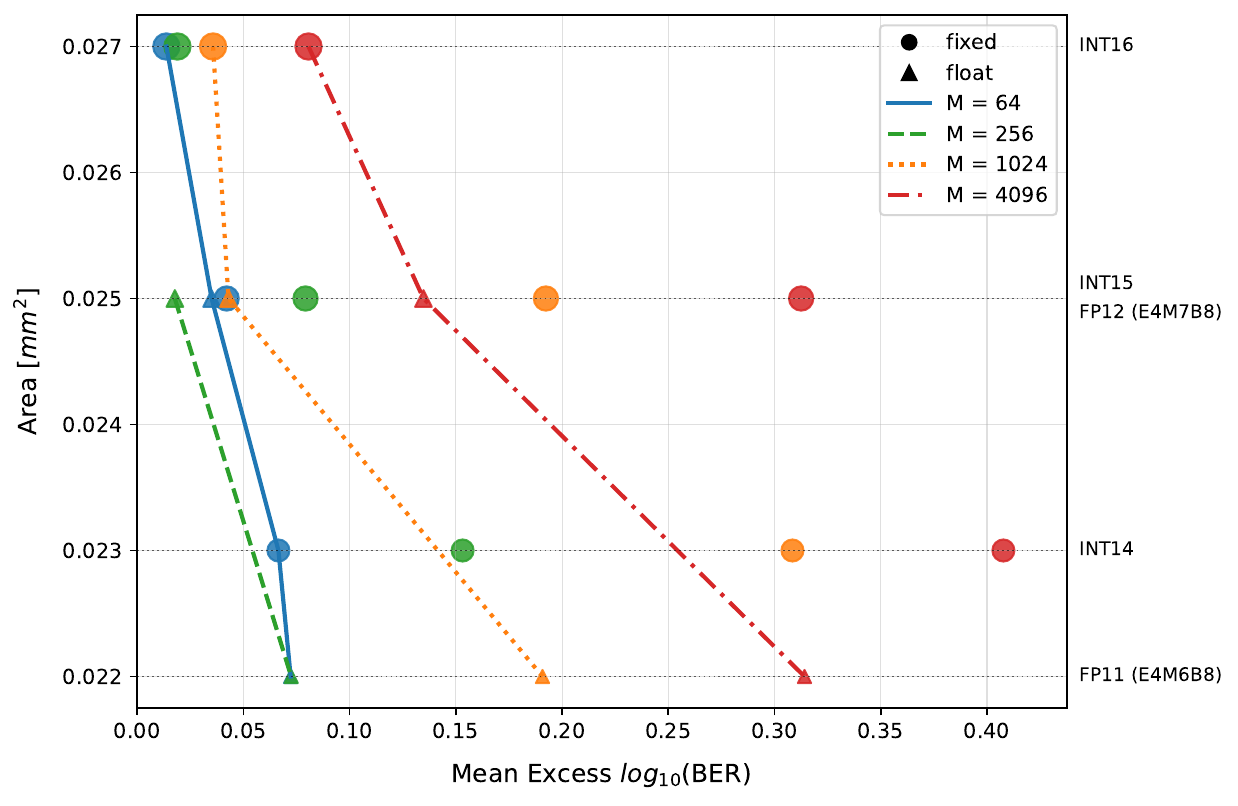}
    \caption{Accuracy--area per-$M$ Pareto fronts.}
    \label{fig:pareto_acc_area}
  \end{subfigure}

  \caption{Joint multi-objective visualization of arithmetic design points for the FFT core. Marker size represents total bit width.
  }
  \label{fig:pareto}
\end{figure*}

Among fixed-point implementations, increasing the fractional precision monotonically improves the numerical robustness, but at the cost of higher power and silicon area. INT16 exhibits the highest power consumption (8.2\,\si{\milli\watt}) and largest area (0.027\,\si{\milli\meter\squared}) among the designs.
In contrast, floating-point formats achieve a more favorable balance between accuracy and hardware cost. FP11 (E4M6B8) reduces power by approximately 19.8\% and area by 12.0\% relative to INT16, while FP12 (E4M7B8) achieves reductions of 17.1\% in power and 7.4\% in area. From an accuracy perspective, FP12 exhibits low mean and bounded worst-case BER degradation across all QAM orders, closely matching the results of INT16.

The advantage of floating-point over fixed-point can be attributed to two structural effects. First, fixed-point formats are uniformly distributed across their value range. In contrast, floating-point formats exploit exponent scaling to cover rare large-magnitude events while preserving mantissa precision where it matters most. Second, when the exponent range is constrained and arithmetic units are simplified, floating-point adders and multipliers can be implemented efficiently.

Fig.~\ref{fig:pareto} provides a complementary design-space view by plotting the per-$M$ Pareto fronts for post-P\&R power and area. For small QAM orders ($M=64$ and $M=256$), multiple configurations cluster near the low-BER, low-power region, indicating that aggressive precision reduction is feasible with minimal system-level impact. As $M$ increases, the Pareto fronts shift to the right, reflecting the growing sensitivity to quantization noise as the QAM constellations tighten.
Across all QAM orders, FP11 and FP12 lie on the Pareto-optimal frontier, offering lower power and area than INT16 at comparable accuracy levels. FP11 trades modest BER degradation for additional efficiency gains compared to FP12. As $M$ increases, the accuracy gap between INT16 and FP11/FP12 widens. 

It is important to note that the floating-point formats considered in this study were fitted with $M=256$ as the operating point. Consequently, their bit width allocation reflects a trade-off between dynamic range and precision optimized for small- to medium-order constellations. Re-fitting the floating-point parameters for higher QAM orders would likely reduce the gap.
This also explains why the mean excess $\log_{10}(\mathrm{BER})$ for $M=256$ is in some cases lower than for $M=64$.

\section{Conclusion}
In this paper, we revisited number formats for classical DSP in view of the recent success of reduced-precision floating-point arithmetic in ML accelerators. We proposed a cross-layer co-design methodology that links (i) application-driven value distributions, (ii) fitting of custom number formats, (iii) arithmetic-unit instantiation, and (iv) joint evaluation of system-level performance and post-P\&R PPA metrics.
As a case study, we investigated a 256-point FFT engine embedded in a \SI{128}{\giga\bit\per\second} optical OFDM transceiver and implemented in a \SI{12}{\nano\meter} FinFET technology. We showed that fitted 11--12-bit custom floating-point formats preserve BER performance close to a 32-bit floating-point reference, while reducing power by up to 19.8\% and silicon area by up to 12.0\% compared to fixed-point designs. The results indicate that reduced-precision floating-point arithmetic can be beneficial for DSP kernels when the format is aligned to the application-level requirements. Future work includes investigating fitting techniques beyond the global assignment, approximate arithmetic, and applying the methodology to other DSP kernels and applications.

\end{document}